\documentclass[
 amsmath,amssymb,
 reprint,
]{revtex4-1}

\usepackage{graphicx}% Include figure files
\graphicspath{{figures/}} %Setting the graphicspath
\usepackage{dcolumn}% Align table columns on decimal point
\usepackage{bm}% bold math

\usepackage[utf8]{inputenc}
\usepackage[T1]{fontenc}
\usepackage{mathptmx}
\usepackage{gensymb}

\begin{document}

\title[A flexible and scalable, fully software-based lock-in amplifier for nonlinear spectroscopy]{A flexible and scalable, fully software-based lock-in amplifier for nonlinear spectroscopy}

\author{D. Uhl}
\author{L. Bruder}
 \email{lukas.bruder@physik.uni-freiburg.de}
\author{F. Stienkemeier}
\affiliation{ 
Institute of Physics, University of Freiburg, 79104 Freiburg, Germany
}

\date{\today}

\begin{abstract}
We demonstrate a cost-effective, fully software-based lock-in amplifier (LIA) implemented on a commercial computer. 
The device is designed for application in nonlinear spectroscopy, such as transient absorption and coherent multidimensional spectroscopy, but may be also used in any other application. 
The performance of our device is compared to a state-of-the-art commercial LIA with nearly identical results for both devices. 
Advantages of our device over commercial hardwired electronic lock-in amplifiers is the improved flexibility in the data analysis and the possibility of arbitrary up-scaling of the number of LIA channels.
\end{abstract}

\maketitle

\section{\label{sec:introduction}Introduction}

In spectroscopy, lock-in amplification is a widely used method to detect weak signals in noisy environments. 
To this end, the signal is modulated at a well-defined frequency and the lock-in amplifier (LIA) efficiently amplifies the signal while suppressing background contribution. 
In particular, in nonlinear spectroscopy, linear background contributions often dominate the weak nonlinear signals and efficient signal recovery methods or background-free detection geometries become essential\,\cite{fuller_experimental_2015}. 
Moreover, some spectroscopy methods rely on optical interference signals, which are prone to phase jitter as an additional noise source\,\cite{scherer_fluorescence-detected_1991}. 

A prominent example, suffering from both experimental conditions, is coherent multidimensional spectroscopy (CMDS) in the optical regime\,\cite{jonas_two-dimensional_2003}. 
Here, numerous background suppression and phase stabilization schemes have been developed\,\cite{fuller_experimental_2015}. 
As a particular efficient scheme, a phase modulation technique was developed, which combines acousto-optical phase modulation with lock-in amplification to extract weak nonlinear signals while removing phase jitter from the signal\,\cite{tekavec_fluorescence-detected_2007}. 
Among the various experimental implementations of CMDS\,\cite{fuller_experimental_2015}, this method offers a remarkable sensitivity which permitted the extension of CMDS to highly dilute samples in the gas phase\,\cite{bruder_coherent_2018} and the detection of extremely weak inter-particle interactions\,\cite{bruder_delocalized_2019}. 
Likewise, the inherent phase jitter correction enabled interferometric spectroscopy at very short wavelengths in the extreme ultraviolet spectral region\,\cite{wituschek_tracking_2020, wituschek_phase_2020}. 
The method has been also combined with a wide variety of detection schemes, ranging from fluorescence\,\cite{tekavec_fluorescence-detected_2007} (including microscopy\,\cite{tiwari_spatially-resolved_2018}), photocurrent\,\cite{nardin_multidimensional_2013}, photoelectron\,\cite{bruder_coherent_2018} and ion-mass detection\,\cite{bruder_phase-modulated_2015, bruder_coherent_2019}. 
A variant of the method was even implemented in the spectroscopy with quantum light\,\cite{lavoie_phase-modulated_2020}.

Commercially available LIAs are based on analog or digital hardwired circuits which poses some general limits on LIA applications, in particular for the aforementioned phase-modulation technique. 
With analog or digital LIAs, the available signal processing tools are predefined on the LIA hardware and cannot be easily extended. 
Likewise, signal processing parameters have to be defined prior to the measurement and cannot be adapted in the post processing which is disadvantageous especially in experiments with long measurement times as common in CMDS. 
Most importantly, many spectroscopy applications would require a large number of LIAs operated simultaneously. 
Examples are spectrograph applications with one-dimensional (1D) and imaging applications using two-dimensional (2D) pixel detectors or time-of-flight ion-mass and electron energy spectrometers. 
While at low resolution, 1D spectrograph data may be still handable with $< 10^2$ LIA channels for which the use of hardwired LIAs is still feasible\,\cite{Xue2015, mao_multi-channel_2015}, in high resolution spectroscopy and for 2D array detectors the number of required LIA channels quickly scales to $\geq 10^3$ which becomes unpractical for electronic LIA circuits. 

These drawbacks of hardware-implemented LIAs can be resolved with software-based LIA schemes. 
In the latter, the signal processing of a LIA is fully implemented as an algorithm running on a central processing unit (CPU) instead of hardwired into electronic circuits, thus reducing the LIA hardware to an analog-to-digital converter (ADC) connected to a computer. 
Since the raw data of the experiment is stored onto the computer's harddrive, flexible post processing with different LIA settings and individual demodulation routines can be applied optimized to the measurement scheme. 
With the speed of state-of-the-art solid-state-drives and the computing power of modern CPUs, it is feasible to scale up the number of lock-in demodulators to $> 10^3$ while keeping the processing time reasonably low. 

Some examples of software-based LIAs have been reported previously which are based on different signal processing concepts\, \cite{augulis_two-dimensional_2011, odonoghue_low_2015, Karki_2013, Chuss_2018, agathangelou2021phasemodulated}. 
One approach uses a digital cavity to extract the phase and amplitude of the signal \cite{Karki_2013}. 
Another one uses Fourier analysis \cite{Chuss_2018}.
In our contribution, we introduce a different type of software-based LIA that directly mimics the circuit of a common, hardwired digital LIA. 
Since this approach provides much more flexibility and universal usage in lock-in detection compared to fully hardware-based concepts, we term our development Universal Lock-In Amplifier (ULIA). 
We demonstrate the performance of the ULIA on the example of phase-modulated wave packet interferometry (WPI) and coherent multidimensional spectroscopy (CMDS).
We compare the algorithm with a state-of-the-art commercial digital LIA and show additional features that come with the use of an algorithm running on a CPU.

\section{\label{sec:experimental_setup}Experimental setup}

\subsection{Lock-in amplification technique}

\begin{figure}
    \includegraphics[width=0.9\columnwidth]{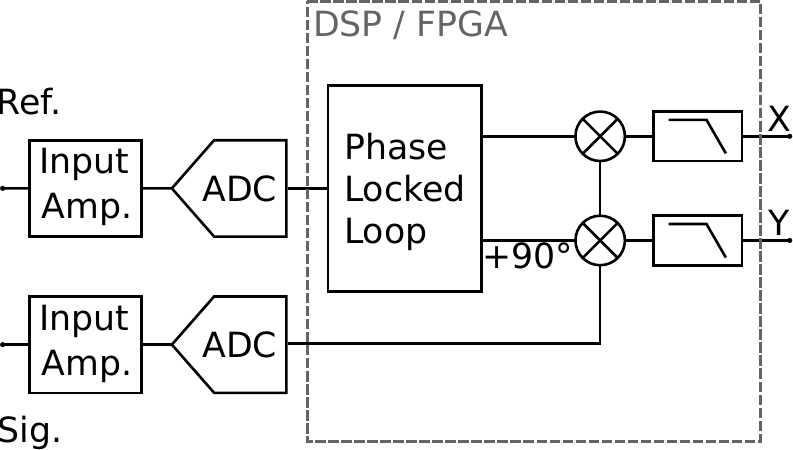}
    \caption{\label{fig1} 
    Basic principle of a common digital LIA. 
    The signal and the reference are pre-amplified by the input amplifiers and digitized by an analog-to-digital converter (ADC) afterwards. 
    The reference signal is synthesized by a phase-locked loop (PLL), split and phase-shifted afterwards to generate two reference signals with a relative phase shift of 90$^\circ$.
    This enables synchronous demodulation of the in-phase (X) and in-quadrature (Y) component of the input signal. 
    To this end, the signal input is multiplied with both reference signals and low-pass filtered afterwards.}
\end{figure}

In Fig.\,\ref{fig1}, we briefly discuss the basic concept of a typical digital LIA. 
This is the common scheme of commercially available LIAs and was chosen as the model for our software-based LIA. 
LIAs are capable of extracting signals from an extremely noisy background, provided the fact, that the signal is modulated at a defined frequency.
To this end, a reference signal of the same frequency is either generated by the device itself or provided by an external source and is usually given as a sine wave. 
For the application discussed here, we focus solely on the case of an external reference signal. 
The LIA multiplies the input signal with the reference, which is commonly called the down-mixing or heterodyne detection.
To isolate the signal at the frequency of interest from the noisy background and all other frequency components, a low-pass filter is applied to the down-mixed signal.
The entire process of mixing and low-pass filtering is called demodulation or phase-sensitive detection.

In the technical implementation of digital LIAs, the signal and the reference are pre-amplified by a programmable input amplifier and then digitized by an ADC. 
The basic steps of signal processing involved in the lock-in detection takes place on a digital signal processor (DSP) or field-programmable gate array (FPGA), as indicated by the grey box in Fig.\,\ref{fig1}. 
The external reference is synthesized by a PLL that extracts its phase, frequency, and amplitude and returns two normalized reference signals differing by a relative phase shift of 90$^\circ$.
The input signal is split up and separately multiplied with the two synthetic reference signals.
The two mixed signals are then passed through a configurable low-pass filter, respectively, which removes the noise from the signal and returns the two outputs $X$ and $Y$, known as the in-phase and quadrature component, respectively.
The amplitude and phase are then easily derived from these values as they can be seen as real and imaginary part of a complex value $Z = X + iY$.
Note that the analysis of a pure sine-wave reference by the PLL enables the ability to do selective measurements at the fundamental frequency and any of its harmonics.

\subsection{\label{sec:ULIA}Implementation of the software-based universal Lock-In Amplifier}

The concept of the developed software-based LIA adapts the scheme of commercial digital LIAs shown in Fig.\,\ref{fig1}.
In our ULIA, the reference and signal is digitized by an ADC (Acqiris, model U5303A, 2 channels, 1 GSa/s, 500\,MHz bandwidth, 12\,bit resolution ). 
With respect to the Nyquist-Shannon sampling theorem it is clear that the sampling rate of the ADC needs to be at least twice the modulation frequency.
For signal modulation at frequencies up to $13\,$kHz, as used in this work, the ADC is well above the critical sampling frequency and one could consider using ADCs with lower sampling rates, which are usually available with higher vertical resolution.
This can be advantageous in case very high dynamic range is needed or data processing is limited by the size of the produced data sets. 
In our application we prefer an ADC with high sampling rate for the possibility to resolve fast transient signals, such as fluorescence decay transients or time-of-flight distributions in photoelectron or ion-mass spectrometry. 
However, for the here presented measurements, a low sampling rate was sufficient as the signal bandwidth is limited by the used pre-amplifier (Femto DLPCA-200), reducing the bandwidth to 500\,kHz. 
Therefore we reduced the sampling rate of the ADC to the lowest possible setting which is 15.625\,MSa/s. 
Data processing is performed on a regular office computer (8\,GB RAM, Intel Core i5-6500, 4 cores at a clock speed of 3.2\,GHz, M.2 sata solid state drive with read/write speed of 500MB/s, operating system Windows 10), which is connected to the ADC by a PCI express interface (2.5GB/s transfer rate).

For the data processing, the ADC data can be either streamed onto the hard drive for post processing at a later stage, or they can be loaded into the RAM for real-time processing with the CPU. 
For most ADCs (including our model) the acquisition and data transfer can only be executed in a sequential manner. 
While the transfer time of the acquired data is optimized for most ADCs and is typically $< 10$\,ms, the time required for storing the data on the hard drive varies considerably depending on the used hard drive model and can take up to ten times the acquisition time.

The ULIA algorithm is implemented in the programming language Python and emulates the three basic components of a LIA, that are the PLL, the signal mixing and the low pass filter.
The signal mixing is a straight-forward operation (multiplication of two arrays). 
Implementing an efficient PLL is slightly more complex and will be discussed in the following.
The task of the PLL is to extract amplitude, frequency and phase of a signal as well as applying an arbitrary phase shift to the synthesized signal (here 0$^\circ$ and 90$^\circ$). 
To realize both tasks, a Hilbert transform\,\cite{2020SciPy-NMeth} is applied followed by a digital biquad filter. 
For the biquad filter, we developed an algorithm which approximates the filter function.
This approximation is done by calculating the difference between the reference input and an internally generated oscillator (sine wave).
This oscillator is then adjusted in order to minimize the phase difference to the reference signal.
The phase-shifted oscillator signal then gives the desired complex-valued normalized reference signal. 
Since the phase and frequency of the oscillator is known, creating a reference at different harmonics of the base frequency is straight-forward.

For the low pass filter a transfer function of a Butterworth filter design\,\cite{2020SciPy-NMeth} is used. 
Readily usable code examples can be found for all common programming languages.
The shape of the low pass filter is defined by the order of the filter and the cutoff frequency $f_{C}$.
$f_{C}$ may be calculated from the filter time constant $\tau$ which is the parameter most of the commercial LIA use.
The relation between these two values is given by,
\[f_{C} = \frac{1}{2 \pi \tau}~.\]
The order of such a filter function can simply be incremented by putting another filter in series.

\subsection{Spectroscopy setup}

\begin{figure}[h]
    \includegraphics[width=0.9\columnwidth]{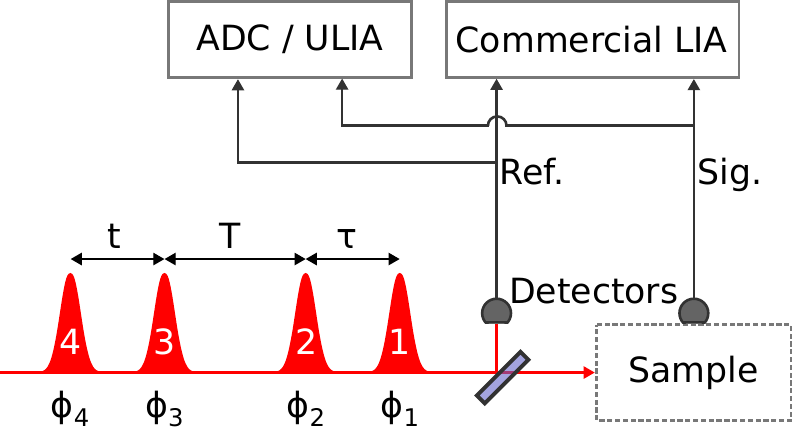}
    \caption{\label{fig2} Experimental scheme. The sample is excited by four phase-modulated laser pulses indicated by the numbers 1-4 with adjustable delays $\tau$, T, and t. The phase $\phi_{i}$ of each pulse is individually modulated, leading to a modulation-beat of the detected signal. A lock-in amplifier is used to demodulate and isolate the nonlinear signal components corresponding to  the rephasing and non-rephasing 2D electronic spectroscopy signals. In parallel to the lock-in amplifier the reference and signal is digitized by an ADC for the ULIA demodulation. In one part of the experiment, only two laser pulses are used to perform linear wave-packet interferometry measurements. }
\end{figure}

A simplified scheme of the experimental setup is shown in Fig.\,\ref{fig2}.
A more detailed description is given in Ref.\,\cite{bruder_coherent_2019}.
In this scheme a four-pulse sequence is generated in a threefold nested optical interferometer.
The inter-pulse delays, here denoted by $\tau$, $t$, and $T$, are controlled by motorized translation stages. 
We perform wave packet interferometry measurements which probe the polarization decay in the sample by exciting the sample with only two pulses and scanning the respective delay between both pulses\,\cite{tekavec_wave_2006}. 
In addition, we conduct two-dimensional electronic spectroscopy (2DES) measurements by applying all four pulses and scan the delays $\tau$ and $t$\,\cite{tekavec_fluorescence-detected_2007}.

In most lock-in detected spectroscopy methods the excitation laser is amplitude modulated.
In contrast, the phase modulation approach uses phase modulation of multiple laser beams exciting the sample in order to induce a modulation of the signal.
In the scope of our experiment, the carrier envelope phase $\phi_{i}$ of each pulse is individually modulated by acousto-optical modulators on a shot-to-shot basis.
With the high repetition rate of the employed laser (200\,kHz) this leads to a quasi continuous modulation of the relative carrier envelope phase between the laser pulses in the kHz regime.
These pulses are used to excite the sample and induce an excited state population, which is probed by fluorescence or ion detection.
Depending on the interaction scheme of the phase-modulated pulses with the sample, the detected signal exhibits a distinct phase signature or quasi-continuous frequency beating according to
\begin{equation}
    \phi_{SIG} = n_{1} \phi_{1} + n_{2} \phi_{2} + n_{3} \phi_{3} + n_{4} \phi_{4} ~,
    \label{eq:gen_condition}
\end{equation}
where $n_i$ denote integer numbers. 

In the experiment, we set the phase beatings to $\phi_{21} = \phi_{2} - \phi_{1} = 5 \mathrm{kHz}$ and $\phi_{43} = \phi_{4} - \phi_{3} = 8 \mathrm{kHz}$. 
Accordingly, the linear wave packet interferometry signals, that are subject to the interaction with only either of the two pulse pairs, are detected at 5 and 8\,kHz, respectively.  
For the nonlinear 2DES experiments, we detect the rephasing (RP) and nonrephasing (NRP) signal contributions at
\begin{subequations}
\begin{eqnarray}
    \phi_{RP} = \phi_{21} - \phi_{43} = 3kHz~, \label{eq:rp}
    \\
    \phi_{NRP} = \phi_{21} + \phi_{43} = 13kHz~. \label{eq:nrp}
\end{eqnarray}
\end{subequations}

While the reference for the linear contributions can be directly extracted from the optical interference of the respective pulse pairs 1, 2 and 3, 4, the nonlinear components need an additional mixing process to generate the respective reference signals for the RP and NRP 2DES signals. 
Furthermore, in order to demodulate both signals in parallel, two commercial LIAs are needed, each for every contribution. 
In previous phase-modulated 2DES experiments, this was done by mixing the linear optical signals with dedicated electronic circuits\,\cite{tekavec_fluorescence-detected_2007, nardin_multidimensional_2013} and separate LIAs for the demodulation. 
Once a measurement is finished, it is not possible to demodulate for an additional signal contribution, e.g. higher-order multiple quantum signals\,\cite{Bruder.2015, bruder_delocalized_2019, Yu.2018b}, without repeating the experiment. 
The ULIA has here the advantage, that the two linear reference signals ($\phi_{21}$ and $\phi_{43}$) are recorded and stored on the computer's hard-drive and arbitrary demodulation processes with any mixing signals can be performed in the post processing. 
The software implementation of respective mixing operations to calculate the higher-order reference signals, requires only straight-forward array multiplication and subsequent band-pass filtering operations.
Compared to the commercial LIA this is a major advantage as it makes it possible to demodulate for any desired phase signature even after the measurement has finished.

\section{\label{sec:results}Results}

\subsection{\label{1D}Application in one-dimensional wave packet interferometry}

To demonstrate the functionality and performance of the developed algorithm, a 1D phase-modulated wave packet interferometry measurement was carried out, comparable to experiments performed in Refs.\,\cite{tekavec_wave_2006, bruder_phase-modulated_2015}. 
As target system we chose an atomic rubidium vapour contained in a spectroscopic cell. 
This system gives us sharp, well defined spectra which is ideal for the analysis of the LIA performance. 
Detected is the fluorescence of the vapor cell for a laser center weavelength of 780\,nm exciting the $\mathrm{5S_{1/2}}\rightarrow \mathrm{5P_{1/2(3/2)}}$ transitions (D-lines).
The signal is split and fed into a commercial LIA (MFLI, Zurich Instruments) and the ULIA for a direct performance comparison of both LIAs. 

At the commercial LIA, the signal and reference input have a vertical resolution of 16\,bit and are sampled at a rate of 60\,MSa/s (signal input) and 30\,MSa/s (reference input), respectively.
The demodulated signal is transferred to the measurement computer via Ethernet connection at a transfer rate of 3.348\,kSa/s, which determines the sampling rate of the demodulated LIA signal.

In comparison, the ULIA features a two-channel ADC with vertical resolution of 12\,bit and sampling rate of 15.625\,MSa/s per channel.
Since demodulation is performed directly on the measurement computer, the ULIA output signal is sampled at the same rate and is not limited by hardware transfer rates as it is the case for the commercial stand-alone LIA.

Data is taken at a laser repetition rate of 200\,kHz and the same demodulation parameters (filter time constant of 30\,ms, filter order of 2) are used for both LIAs.
At each inter-pulse delay, the demodulated data of the commercial LIA is averaged over 300\,ms amounting to 1004 averaged sample points per inter-pulse delay. 
For the ULIA, we average the demodulated data over 240\,ms, which corresponds to 4.6875\,MSa. 
The reduced averaging interval for the ULIA is chosen to simplify the synchronization of the data acquisition between the two LIA devices. 
For the chosen averaging time, the raw (not averaged) data for signal and reference amounts to a data size of 18\,MB per delay point. 

\begin{figure}
    \includegraphics[width=0.9\columnwidth]{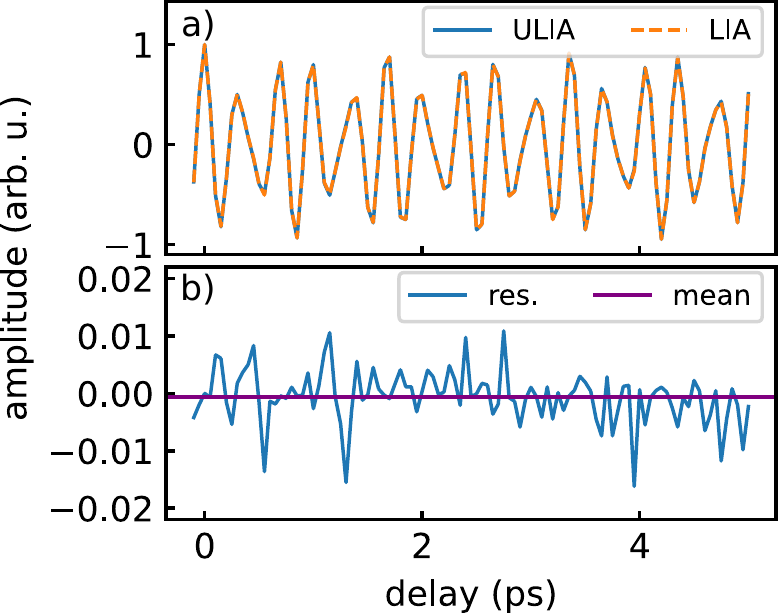}
    \caption{\label{fig3} 1D wave packet interferometry measurement comparing the performance of the ULIA with a commercial LIA. 
    (A) shows the demodulated in-phase signal of both methods with ULIA data (blue) and commercial LIA data (orange). 
    (B) Residual plot of the data sets from (A) (blue) and its mean value (purple).}
\end{figure}

Fig.\,\ref{fig3} shows the recorded time domain interferograms.
(a) shows the real part of both demodulated signals, which are normalized to provide a better comparison.
This is reasonable since we focus our analysis on the SNR of the data sets. 
Without normalization of the LIAs outputs, a small difference of the amplitude ($<\mathrm{1\,\%}$) is observable (not shown).
Exemplary, for the D1 line, the commercial LIA yields an absolute amplitude of $\mathrm{6.86 \cdot 10^{-4}\,V}$ and the ULIA an amplitude of $\mathrm{6.82 \cdot 10^{-4}\,V}$.
This small difference might originate from different damping factors in the signal cables and calibration uncertainties of the different ADC used in both LIAs.
A qualitative analysis of the plotted data in Fig.\,\ref{fig3} shows that both demodulation methods return a nearly identical result.
To quantify the congruence of both data sets, the residuals are shown in Fig.\,\ref{fig3}b.
The subtracted data results in a white noise which averages around zero with a standard deviation of $\mathrm{7 \cdot 10^{-3}}$, or 0.7\,\%, respectively.

\begin{figure}
    \includegraphics[width=0.9\columnwidth]{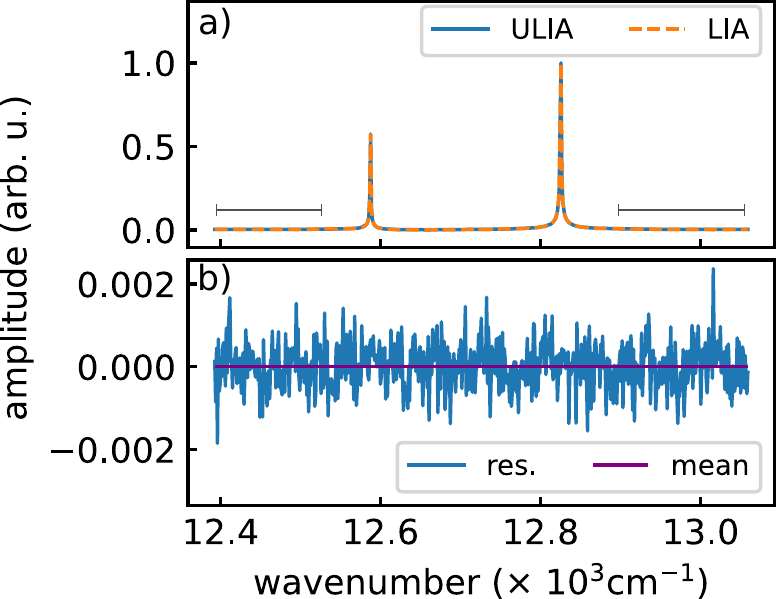}
    \caption{\label{fig4} (a) Fourier spectra of the WPI measurement, from ULIA data (blue) and commercial LIA (orange). 
    (b) Residues of the data sets from (a) (blue) and its mean value (purple).}
\end{figure}

A Fourier transform of the time domain data yields the spectrum of the excited transitions in our sample.
In Fig.\,\ref{fig4} the absolute value of both spectra are shown.
Note that the spectra are normalized to their highest peak for a better comparison.
As in the case of the time domain data, the spectra also match nearly perfectly.
Fig.\,\ref{fig4}b shows the residuals reflecting a white noise spectrum with zero offset and a standard deviation of $\mathrm{5.5 \cdot 10^{-4}}$.
Disregarding the constant scaling factor between the two LIA outputs, these results imply that the spectroscopic data (peak positions and shape) match to a high degree.

\subsection{Performance comparison between the LIA methods}

To further compare the two methods, the noise level in Fig.\,\ref{fig4}a is evaluated.
To this end, a subset of the data (indicated by grey bars in Fig.\,\ref{fig4}a is evaluated, yielding a standard deviation of $\mathrm{2.5 \cdot 10^{-4}}$ for the ULIA and $\mathrm{5.9 \cdot 10^{-4}}$ for the commercial LIA.
Hence, in this experiment, the ULIA results in a lower noise level by a factor of 2.4 compared to the commercial LIA and thus in a better SNR.

The difference in SNR is attributed to the different ADC bandwidths in the devices in combination with the pulsed signals that are digitized.
The ADC bandwidth of the ULIA is 7.8125\,MHz.
For the commercial LIA it is $ \leqslant $500\,kHz.
The latter is in principle already sufficient to adequately sample the applied modulation frequency of 5\,kHz in the experiment.
However, the fast fluorescence detector (amplified photo multiplier tube; Hamamatsu H10720-20) returns a short fluorescence spike (width on the order of 1\,$\mu s$) for each laser excitation.
These signal pulses are fully resolved by the ADC of the ULIA but not by the one in the commercial LIA.
Essentially, the LIAs integrate over these signal spikes and extract the slow 5\,kHz-modulation from the smoothed signal.
Hence, insufficient sampling of the signal spikes adds fluctuations on the integrals which can increase the noise in the demodulated output signal.
This issue is independent of the detected observable and may arise whenever fast, single-shot detection is used, as it is common in many experimental schemes, e.g. in single-counting photon or electron detectors required in low signal conditions.
The effect, however, may be equalized by smoothing the detector signal with suitable integration electronics prior to feeding the signal into the LIAs.
Furthermore, they may also depend on the noise characteristics and dynamic range of the used ADC.
The latter strongly depends on adequate matching of signal amplitude with the range of the ADC's input amplifier (see section C). 
Note, that the reference signal is acquired with a slow photodetector which avoids sampling problems in its digitization. 

One advantage of commercial LIAs is that they provide a demodulation and signal analysis in real time as the raw data streams through the DSP/FPGA.
This is not the case for the ULIA with the algorithm running on a CPU, where data processing is conceptually performed in a sequential manner, that is, the data acquisition and storage onto the hard drive followed by the data demodulation. 
Both processes strongly depend on the computer hardware. 
Due to the architecture of most ADCs, real time data streaming onto the computer's hard drive is not possible and the data can only be transferred in data blocks after acquisition has completed. 
As outlined above, a data block of 18\,Mb has to be transferred from the ADC to the computer at each inter-pulse delay. 
Transferring the data from the ADC to the measurement computer takes about 8\,ms and 18\,ms to write the data onto the hard drive.
The computing time for the demodulation of one data set is $\approx 110\,$ms which adds up to $\approx 136\,$ms for the whole data processing (data acquisition, data transfer and data demodulation) at each delay step.
This is well below the integration time of 300\,ms for each delay step and can thus be done in parallel of the acquisition of the next delay step.
This permits real-time lock-in demodulation with a minimal time lag of ~418\,ms.

In view of real-time demodulation of the acquired signals, distribution of the workload over multiple CPUs is a common strategy to speed-up data processing. 
This is partially possible for other software-based LIA concepts, like the digital cavity\,\cite{Karki_2013}. 
In contrast, in the ULIA scheme, the subroutines of the signal demodulation cannot be parallelized due to the used transfer functions which are mostly recursive. 
While this accounts for the demodulation of one signal channel, in many spectroscopy applications more than one demodulation channel is needed in order to capture all signal components. 
Examples are the demodulation of 1D and 2D array detectors\,\cite{Xue2015, mao_multi-channel_2015} or the parallel detection of linear and higher-order signals\,\cite{Bruder.2015, bruder_delocalized_2019, Yu:19}.
Modern processors for personal computers reach up to 64 CPU cores, which can be utilized as 64 demodulators for an on-line usage. 
Compared to the commercial devices this is a cost effective replacement in a lot of application fields for LIAs, even beyond spectroscopy applications. 

From this we conclude that for applications where real-time demodulation is indispensable, it is advisable to use a regular LIA offering fast processing times and real-time data demodulation with minimum time lag. 
If, however, demodulation of many signal channels is required, scaling the number of commercial LIAs may be inappropriate and the ULIA combined with many CPU cores offers an alternative fast demodulation method.

In case real-time data demodulation is not of highest priority, the ULIA opens-up additional possibilities not available in standalone LIAs offering only real-time data processing and no raw data storage. 
With the ULIA it is possible to first stream the complete measurement onto the hard drive and demodulate the data afterwards, e.g.\ with various different LIA settings to find optimum performance parameters.
In this case, the loading time of the data from the hard drive adds to the overall computing time for the data demodulation.
With our setup the reading time is nearly equal to the writing time which is about 18\,ms per 10\,Mb-data-block.
However, the reading and writing time varies strongly with the used hard drive, from 2\,ms up to over 1\,s in our experience. 
Hence, it is advisable for such tasks to use hard drives with fast writing and reading times. 

Demodulating data in the post processing also opens new fields of applications for lock-in detection.
We have recently successfully implemented the ULIA algorithm with array-type detectors where $\geq 1000$ demodulators are required to demodulate each data pixel. 
This is subject of an upcoming publication. 

\subsection{\label{sec:2D}Application in coherent multidimensional spectroscopy}

\begin{figure*}
    \includegraphics[width=1.\textwidth]{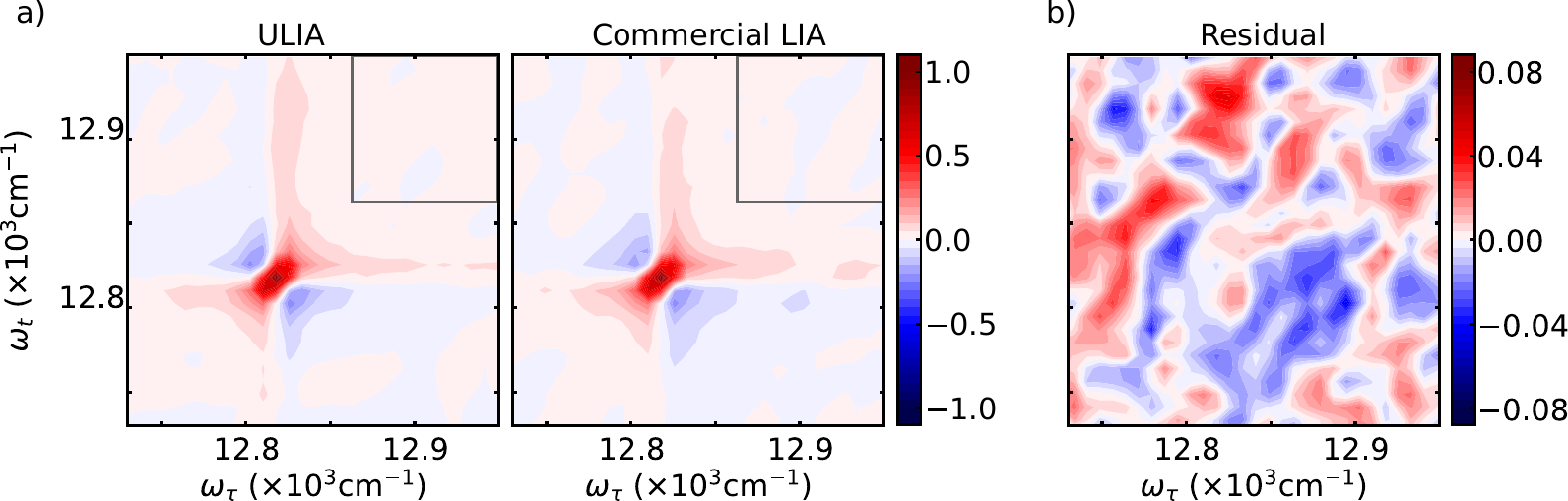}
    \caption{\label{fig5} Normalized 2DES spectra of the D2 line in atomic Rb.
        a) Real part of the measured rephasing spectra for demodulation wiht the ULIA (left) and the commercial LIA (right). 
        The grey box indicates the data used for evaluation of the noise floor. 
        b) Residual plot of the spectra in (a).
        Note that the color scale is enhanced.}
\end{figure*}

As a specific example of a real-time multi-demodulation application we demonstrate 2DES measurements of the Rb vapor sample using the ULIA. 
This experiment utilizes the complete four-pulse train from the three nested Mach–Zehnder interferometer scheme as described above. 
The nonlinear interaction of the pulse train with the sample leads to RP and NRP 2DES signals at the beat frequencies $\phi_\mathrm{RP}=3\,$kHz and $\phi_\mathrm{RP}=13\,$kHz, respectively. 
Hence, two separate LIA demodulation channels are needed to extract both signals from the data simultaneously. 

Moreover, an adequate reference signal has to be computed for both signals. 
The reference signal is retrieved from the linear optical interference of the four-pulse train. 
This interference signal contains the pair-wise beatings between the four laser fields, however, no nonlinear mixing contributions. 
Thus, the reference signal for the demodulation of the nonlinear 2DES signal components has to be computed from the linear beat signals $\mathrm{\phi_{21}=5~kHz}$ and $\mathrm{\phi_{43}=8~kHz}$ as $\phi_\mathrm{ref}=\phi_{21}\pm \phi_{43}$. 
For the signal demodulation with the commercial LIA, a frequency mixer programmed onto a DSP is used to generate the sum ($\mathrm{13~kHz}$) and difference frequency ($\mathrm{3~kHz}$) components of the linear beat signals which are fed into two separate demodulators of the commercial LIA. 
The procedure is simpler for the ULIA. 
Here, the linear interference signal is directly digitized. 
Based on the code blocks described in Sec.\,\ref{sec:ULIA} we constructed band-pass filters to extract the $\mathrm{\phi_{21}=5~kHz}$ and the $\mathrm{\phi_{43}=8~kHz}$ beat components from which we computed the sum/difference frequency signals according to
\begin{equation}
{\displaystyle \sin(\phi_{43} \pm \phi_{21} )=\sin \phi_{43} \cos \phi_{21} \pm \cos \phi_{43} \sin \phi_{21} }\, .
\end{equation}
The resulting $\mathrm{13~kHz}$ and $\mathrm{3~kHz}$ signals are then fed into two separate PLLs to generate the respective reference signals for the demodulation. 

The measured 2D correlation spectra (real part of the sum of RP and NRP spectra) are shown in Fig.\,\ref{fig5}a.
Both graphs show the same measurement using the two different LIA devices, left using the ULIA and right using the commercial LIA.
For a better comparison of the results, the spectra are zoomed on the $\mathrm{Rb} ~ 5S_{1/2} \rightarrow 5P_{3/2}$ resonance (D2 line) and their amplitude is normalized to unity.

These 2D spectroscopy measurements provide another benchmark for the comparison between commercial and our software-based LIA. 
On the one hand, the SNR conditions in the nonlinear spectroscopy experiment are much more demanding than in the 1D spectroscopy above.  
Hence, the signal recovery capabilities of the LIAs are more challenged in this application. 
On the other hand, the 2D spectra offer more spectroscopic parameters to compare between the two methods, e.g. 2D line-shapes and positions. 
Furthermore, the detection of nonlinear signals requires an additional frequency-mixing step to prepare the reference signal which is not the case for linear spectroscopy. 
As outlined above, the two LIA methods differ in this signal processing step. 

As with the 1D measurement above the 2D spectroscopy results from both LIAs are nearly the same in terms of size, shape and position of the signal response.
To quantify the differences between the two data sets, their residuals are plotted in Fig.\,\ref{fig5}b.
The residual plot shows the pattern of white noise with a standard deviation of 0.041, confirming that the two demodulation methods yield the same spectroscopic information.
Furthermore, we analyze the noise floor of both 2D spectra from Fig.\,\ref{fig5}a. 
To this end, we compute the standard deviation of the data inside the grey box of Fig.\,\ref{fig5}a.
For the normalized data sets, the inverse of the standard deviation is a good approximation for the SNR of the measurement.
We get a slightly better SNR value for the ULIA (standard deviation of 0.025), compared to the commercial LIA (standard deviation of 0.036). 
At the given experimental conditions, we assume that the dominating noise source are the ADCs, which introduce uncorrelated noise in the data sets of Fig.\,\ref{fig5}a.
With this assumption one expects a standard deviation of the residuals of 0.044, which explains the large standard deviation of 0.041 in the residual plot of Fig.\,\ref{fig5}b.
This confirms our assumption that the demodulation of both methods, ULIA and commercial LIA, strongly depends on the quality of the digitized raw data.

\section{Discussion}

In our work, we compared two LIA devices, i.e., a commercial state-of-the-art digital LIA and a software-based custom device. 
In principle, both devices are based on a software-implemented algorithm. 
The commercial LIA used here performs the signal processing with an algorithm implemented on a DSP instead of using a fully hardwired circuit. 
In contrast, the ULIA runs the algorithm on a CPU. 
While the DSP is optimized for digital signal processing, the CPU solution provides simple user access to the algorithm and enables flexible modifications to optimize the measurement task for the specific user experiment. 

By directly comparing both devices, we found quantitatively the same performance for both devices. 
Minor differences in the $< 1\,\%$ range are attributed to the different ADCs used in both devices.
In fact, the pre-amplification and digitization of the signal show the greatest effect on the SNR performance of the ULIA and thus offer room for further improvement. 
The bandwidth of the used ADC should cover the relevant frequency spectrum of the experiment and its vertical resolution directly affects the achievable dynamic range of the LIA device. 
Hence, the latter parameter usually has the strongest impact on the signal quality. 
Since in many applications, modulation frequencies are $< 50\,$kHz, commercial sound cards with 24\,bit resolution and 192\,kSa/s could offer an appealing cost-effective alternative as ADCs, offering higher dynamic range than commercial LIAs. 
This, however, will require appropriate signal smoothing prior to the ADC to ensure sufficient digitization bandwidth as implied in our study. 
Likewise, in many laboratories, the detection electronics include an ADC which is already optimized to the requirements of the experiment. 
In this case, adding the ULIA to the experiment would only require the implementation of the source code described in this paper. 
Especially for application where a large scale of LIAs is mandatory, the costs could become a substantial issue compared to the hardware needed for applying our algorithm.

In ultrafast spectroscopy, signals are naturally modulated by the repetition rate of the pulsed femtosecond laser sources. 
In many experiments this feature is combined with switching experimental conditions for background subtraction. 
This, in principle, can also be viewed as a shot-to-shot lock-in demodulation where apart from a software LIA\,\cite{augulis_two-dimensional_2011} much simpler algorithms subtract the signals on a shot-to-shot basis suffice. 
With the here implemented phase modulation approach shot-to-shot subtraction methods are not an option as cycling of many phase positions is needed. 
Furthermore, the signal is modulated by two frequencies. 
In this case, and also in combination with other modulation or subtraction techniques, the LIA can simply separate the individual signal contribution, as long as the individual modulation frequencies significantly differ. 
Related to this, the SNR performance in phase modulation spectroscopy should improve with increasing modulation frequency and several groups have developed setups offering modulation frequencies $> 100$\,kHz\,\cite{karki_phase-synchronous_2016, autry_single-scan_2019}. 
In our experience, however, the SNR advantage saturates for modulation frequencies $\gtrsim 1\,$kHz. 
We attribute this to 1/f noise which usually dominates in the range $< 1\,$kHz and converges to a white noise of roughly constant level at larger frequencies\,\cite{Kearns:17}. 
In case the experiment can be operated only at low repetition rates ($\ll 1\,$kHz), phase synchronous undersampling can be implemented to enhance the SNR\,\cite{bruder_phase-synchronous_2018}. 

A major asset of the ULIA is its flexibility in the signal processing routines. 
While commercial LIAs offer demodulation on a single frequency and its harmonics, the ULIA offers demodulation on arbitrary nonlinear mixing frequencies. 
This simplifies for instance the detection of rephasing and non-rephasing signals in 2DES and may be further exploited in higher-order nonlinear spectroscopy. 
 
In contrast to standalone LIAs, the ULIA streams the raw data of an experiment on the computer's hard drive, making it available for advanced post processing. 
Custom filter algorithms can be implemented optimized to the specific measurement task and the number of demodulators can be realistically scaled to the $10^3$ range. 
On the contrary, streaming the data on to a hard drive slows down the real-time demodulation of the experimental data. 
Still, in most cases real-time analysis with time lags $< 1\,$s are feasible with our approach. 
Nonetheless, in certain applications, e.g. in the optimization, characterization or trouble shooting of an experimental setup, real-time analysis with negligibly small time lags are pivotal due to which a commercial LIA might be indispensable for such tasks.

\section{Conclusion}

In conclusion, we have developed an efficient software-based LIA, termed ULIA, consisting of an ADC and an algorithm running on a CPU of a regular office computer. 
The core of the algorithm is a phase-locked loop, which is implemented as the combination of a Hilbert transform and an approximated biquad filter. 
The focus of our development lies on applications in spectroscopy, where LIAs are used to extract spectroscopic signals with high dynamic range. 
The newly developed LIA was applied in a linear wave packet interferometry and a nonlinear 2DES experiment under challenging SNR conditions. 
We found a very similar signal-to-noise performance compared to a commercial state-of-the-art LIA confirming the efficiency and reliability of the ULIA. 
The high flexibility of the fully software-based ULIA, including demodulation at arbitrary nonlinear signal combinations and facile up-scaling of parallel processed detection channels, opens-up many spectroscopic applications and the extension of the phase modulation technique towards new detection types. 

\section*{Data availability}
An installable module and the source code of the here presented algorithm is available on the ULIA Python Package Index (https://pypi.org/project/ulia) and can be used with appropriate citation of this work.

Data presented in this paper may be obtained from the authors upon reasonable request.

\section*{Acknowledgement}
Funding by the Bundesministerium für Bildung und Forschung (BMBF) STAR (05K19VF3), by the
European Research Council (ERC) Advanced Grant COCONIS (694965), and by the Deutsche
Forschungsgemeinschaft (DFG) GRK 2079 is gratefully acknowledged.

\section*{Peer-Review}
This article has been published in \textbf{Review of Scientific Instruments} (\url{https://aip.scitation.org/doi/10.1063/5.0059740}).

\bibliographystyle{apsrev4-2}
\bibliography{ULIA_paper0}

\end{document}